# Formation Flight Control of Multi-UAV System Using Neighbor-based Trajectory Generation Topology

BATOOL IBRAHIM, HASAN NOURA
Faculty of Engineering
Lebanese University (LU), Islamic University of Lebanon (IUL)
Beirut
LEBANON
batoolaibrahim@hotmail.com, hassan.noura@iul.edu.lb

*Abstract:* In this paper, a distributed formation flight control topology for Leader-Follower formation structure is presented. Such topology depends in the first place on online generation of the trajectories that should be followed by the agents in the formation. The trajectory of each agent is planned during execution depending on its neighbors and considering that the desired reference trajectory is only given to the leader. Simulation using MATLAB/SIMULINK is done on a formation of quadrotor UAVs to illustrate the proposed method. The topology shows very good results in achieving the formation and following the reference trajectory.

## 1 Introduction

One of the hottest topics nowadays is Unmanned Aerial Vehicles (UAVs). UAVs are powered flying vehicles that do not carry a human operator, can be operated remotely or autonomously, and can carry a payload [1]. The usage of a single small scale UAV in both military and civilian applications is already a reality in today's market. They are promising and useful in many applications such as payloads transportation [2], targets searching [3], in addition to educational purposes. However, UAVs are still not used extensively due to some various limitations imposed when using a single UAV. Such limitations include load tolerance, searching objects in large area, battery life, etc. Motivated by these potential applications, and to overcome most of these limitations, researchers are more and more attracted to the use of multi-UAV systems, or in other words, formation flights.

Formation flight is the coordinated flight of UAVs towards achieving a certain goal. This is done through the use of a formation controller which works on coupling the UAVs' states to accomplish the desired formation. In the literature today, there are three main types of formation control structures or configurations: Leader-Follower, Virtual, and Behavior-based. In Leader-Follower structure, one agent is designated as a leader while others are treated as followers [1]. The formation flight mission trajectory is loaded in the leader, and the followers are tasked to track it in order to achieve and maintain the formation. Works on such structure exist nowadays, for example, having the agents fly in a specific shape form, such as the work in [2], or having multiple leaders like in [3]. Accordingly, Leader-Follower formation is desirable and attractive due to its simplicity [1] and its wide implementation in multi-agent formation [4], but, it is poorly robust to the leader's failure [5]. In virtual structure, the formation is treated as a single geometric entity [6]. Each agent in the formation has its own trajectory to follow, where the group of all trajectories form the desired formation [7]. Examples of experimental works of such control strategy can be found in [8] and [9]. This structure solves the problems of reliability and error propagation presented in the Leader-Follower structure. However, the computational and communication cost during the formation control increases [7]. Finally, in Behavioral-based structure, the formation control task is broken down into smaller tasks referred to as behaviors [10]. These behaviors are then weighted based on priority levels and thus averaged in order to generate the formation control actions [11]. This structure is easily self-organized and scalable [7], but, it is hard to guarantee that the formation has converged to the desired configuration [6]. Such structure is found in the work of [12] and [13]. However, in this paper, Leader-Follower structure is taken into account. Moreover, in formation control, four types of architectures can be distinguished in the literature today: Centralized, Decentralized, Distributed, and Hierarchal. In centralized architecture, all the agents in the formation are controlled by one centralized controller where all computations and processing are carried out. Several works can be found on such architecture like in [14] and [15]. Such control architecture generally offer the most robust and optimal path planning solutions [16], but, on the other hand, it has a weak robustness with respect to the fault of the central processor [1]. In decentralized control system, the overall system is not controlled by a single controller, but by several independent controllers [1]. This architecture exists in works in the literature today, such as the work in [17]. Since the controller in the decentralized control architecture has no knowledge about the states of the other subsystems, this architecture could not be applied for autonomous flight formation control [6]. However, the wide use of name "Decentralized control" in some formation control literature, for instance the works of [18] and [19], could be conflicted with the distributed architecture. The distributed control architecture evolves from the decentralized control, but with sharing local information [1], where it uses information from neighbors, such as the works in [20], [21], and [22]. Distributed control strategy are easily scalable, however, communication faults and delays may take place. Finally, in Hierarchal control architecture, different agents are assigned different levels of autonomy and decision making strengths [16]. Tasks are distributed among agents based on levels of autonomy. Upon comparison, hierarchal controllers are less scalable, more optimal and require more communication links than decentralized and distributed controllers. However, they are more scalable, less optimal and require less communication links than centralized controllers [16]. In this work, the distributed formation architecture will be used because of the need of data sharing and communication between the formation's agents in the proposed topology, in addition that such formation architecture is the most used and successful one for autonomous flight formation control [6].

Several works and methods are proposed to control a formation flight in the literature nowadays. Some methods are based on optimization approaches, such as the work in [7]. Other methods propose the use of consensus protocol which aims at achieving an agreement of all agents in a multi-agent system in order to reach a predefined fleet goal [10], like in [22]. However, several researchers have taken into account formation control using trajectory generation or planning. For example, in [24], according to the characteristic of radio communication path loss constraints, the trajectory planning method based on distributed model predictive control is proposed. In [25], a three-dimensional path planning for UAVs using fast marching method is provided for generating obstacle-free paths by adjusting parameters. Moreover, in [26], trajectory planning algorithm for fixed wing communication relay UAVs in urban environment is presented by considering dynamic constraints. However, the main contribution of this paper is proposing and applying new formation control technique, which is the neighbor-based trajectory generation topology. Such method can be considered as a combination between the concept of trajectory generation in the first place, and the concept of neighbor-based data used in the consensus techniques in the second place. In the conventional consensus methods, while using neighbor-based information to control a formation, an assumption on the dynamics of the system is taken into account, such as considering a dynamics of a double integrator for each agent in the formation like the work in [22]. However, this paper introduces the neighbor-based concept while generating the desired trajectory for each agent in the formation without any need to consider assumptions on the dynamics of the agents. Accordingly, the non-linearity of the system can still be presented while controlling the system and thus making such proposed method more realistic. Moreover, generating trajectories for the agents in the formation using neighbor-based data in Leader-Follower formations was not introduced or applied before. This paper proposes and applies such topology on a formation of quadrotor UAVs with considering different scenarios.

The remaining part of this paper is organized as follows. In Section 2, the general concept of trajectory generation is presented in addition to discussing the proposed trajectory generation method for formation control. In Section 3, the control of a single quadrotor will be considered in order to be able to introduce the formation controller. Section 4 is devoted to present and discuss the simulation results using MATLAB/SIMULINK. Finally, conclusion is made in Section 5.

## 2 Trajectory Generation

Trajectory generation consists of defining a set of paths parameterized by time in order to accomplish a specific mission, with of course taking into account the dynamic constraints of the system agents and the surrounding environment [23]. The importance and complexity of such problem is reflected in the number of research activities about this field in the literature today, especially in formation control. Accordingly, trajectory planning or generation can provide a trajectory for the UAVs in the formation to successfully complete the flight mission. Depending on the objectives of the mission, the generation of the trajectory can be done online or offline [23]. Offline generation is done prior to the execution of a mission and is not altered once the mission is underway. Using such mode alone is usually not sufficient to successfully fly a UAV or a formation of them along a desired path, where they usually fly in environments unaccounted for disturbances that would lead to a loss of time optimality and an increase of accelerations and velocities beyond the safe limits. Also, generating a trajectory initially will not allow to take advantage of changing the path, the desired shape, and the configuration of a formation online. Therefore, in such cases, online trajectory generation is preferably used. By implementing such trajectory generation method, there is an ability to change the formation configuration online. In some cases, researchers opted to use offline path generation with the incorporation of online trajectory generation to enhance the accuracy of path tracking [16]. The work in this paper takes advantage of the online trajectory generation to control a formation of quadrotors.

The set $V = \{1, 2 \ldots, n\}$ includes the indices of the UAVs from 1 to n, where n is the number of agents in the formation. The leader's index set $V_L = \{i \in V : \text{agent i is a leader}\}$. The objective of the formation controller is to guarantee that all UAVs follow the formation reference trajectory with some constant biases $d_{i0} = [d_{Xi0}, d_{Yi0}]$, so that the agents keep a constant formation pattern. Note that this work considers the formation translational dynamics. The altitudes of all UAVs are assumed to be stabilized and constant, such that $Z_i = constant$.

$x_i = [X_i, Y_i]^T$ presents the planar position of UAV$_i$. The reference formation trajectory (RFT) is presented by $r(t) = [r_x(t), r_y(t)]^T$. Therefore, the Leader-Follower formation control is achieved if for each UAV$_i$ and for some initial conditions $x_i(0)$ where i=1…, n:

$$\lim_{t \to \infty} ||x_i - r(t) - d_{i0}|| = 0 \quad (1)$$

Therefore, the desired position of UAV$_i$ evolves according to $x_i^d(t) - r(t) = d_{i0}$, which leads to:

$$x_i^d(t) = r(t) + d_{i0}$$
$$\dot{x}_i^d(t) = \dot{r}(t) \quad (2)$$

Therefore, the formation task contains two parts. The first part is the desired trajectory RFT, and the second one is the biases away from the RFT. But, since Leader-Follower formation structure is taken into account, the formation flight mission trajectory is only loaded in the leaders. So, the desired trajectory $x_i^d(t)$ for UAVi is not available and cannot be used in the formation controller design for the followers. However, neighbor-to-neighbor information exchange is available because of the use of Distributed control architecture.

The neighbor set of a given agent i contains the UAVs with an inter distance with UAV$_i$ smaller than d, where d is a positive scalar. Therefore, the neighbor set of UAVi is represented by:

$$N_i = \left\{ j \in V : \sqrt{(X_j - X_i)^2 + (Y_j - Y_i)^2} \leq d \right\} \quad (3)$$

Therefore, the formation problem is how to generate the trajectories for the UAVs in order to attain the formation task using only the neighbors' states.
The sum of the relative position state vectors:

$$\sum_{j \in N_i}(x_i - x_j - d_{ij}) \quad if \; i \in V - V_l$$

$$\sum_{j \in N_i}(x_i - x_j - d_{ij}) + x_i - r - d_{i0} \quad if \; i \in V_l$$

The inter-distance is given by $d_{ij} = d_{i0} - d_{j0}$. Then, the summation equations can be rewritten as follows:

$$\sum_{j \in N_i}(x_i - r - d_{i0} - (x_j - r - d_{j0})) \quad if \; i \in V - V_l$$

$$\sum_{j \in N_i}(x_i - r - d_{i0} - (x_j - r - d_{j0})) + x_i - r - d_{i0}$$
$$if \; i \in V_l$$

Thus, the available desired trajectory or the generated trajectory can be introduced for each UAV$_i$ as follows:

$$\bar{x}_i^d = \begin{cases} \frac{1}{|N_i|}\sum_{j \in Ni}(x_j + d_{ij}) & if \; i \in V - V_l \\ \\ \frac{1}{|N_i+1|}(\sum_{j \in Ni}(x_j + d_{ij}) + r + d_{io}) & if \; i \in V_l \end{cases} \quad (4)$$

Therefore, the trajectory of each agent is generated online depending on the surrounding neighbors' positions in the first place. For each UAV$_i$, it is enough to have the measurements of the relative positions with respect to its neighbors instead of the need for the desired reference trajectory for all agents as shown in (2), thus, the desired biases from the reference will be achieved automatically. However, if UAV$_i$ is a leader, besides the relative measurements, it also needs to obtain the relative position with respect to the reference formation trajectory (RFT) in order to introduce it to the formation.

It is not necessary to take the same desired biases or inter-distances at all the flight time. There exists an ability to change them while online generating the trajectory, depending on the required application.

## 3 Problem Formulation

In this work, each agent in the formation is considered as a quadrotor, which is a rotary UAV with four motors. However, the proposed formation control strategy can be applied for different types of agents.

The dynamics of a quadrotor are modeled as the motion of rigid body in 3-D space under a thrust force and three moments.

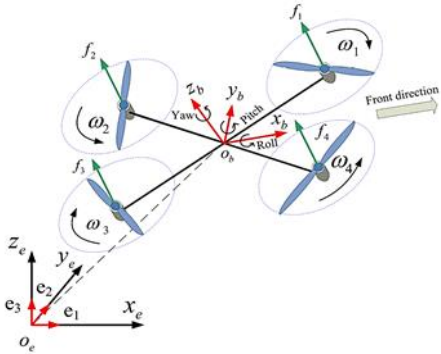

Fig.1 Quadrotor scheme. The inputs are four thrust forces generated by the four propellers. The attitude is represented by the Euler angles.

As Euler angles representation is used, the state of quadrotors, as shown in Figure 1, is represented in an inertial frame $o_e x_e y_e$ and a body-fixed frame $o_b x_b y_b$. The unit directional vectors of the inertial reference frame is denoted by $\{e_1, e_2, e_3\}$, while the unit directional vectors of the body-fixed frame by $\{b_1, b_2, b_3\}$.

The structure of the quadrotor is symmetric in the plane $o_b x_b y_b$. The dynamics of the rotors and propellers are not considered. It is assumed that the thrust of each propeller is directly controlled.

For each propeller, the thrust and the torque depend on its rotating velocity with respect to the coefficients $K_T$ and $K_\tau$. Then, the rotating velocities of the four propellers $\omega_1, \omega_2, \omega_3$, and $\omega_4$ are related to the total thrust $F_T$, which is the summation of the four rotors' thrusts ($F_T = f_1 + f_2 + f_3 + f_4$), and the three moments $\tau_\theta, \tau_\varphi$, and $\tau_\psi$ by the Distribution matrix, according to [22], as follows:

$$\begin{bmatrix} F_T \\ \tau_\varphi \\ \tau_\theta \\ \tau_\psi \end{bmatrix} = \begin{bmatrix} K_T & K_T & K_T & K_T \\ K_T l_a & K_T l_a & -K_T l_a & -K_T l_a \\ -K_T l_a & K_T l_a & K_T l_a & -K_T l_a \\ K_\tau & -K_\tau & K_\tau & -K_\tau \end{bmatrix} \quad (5)$$

Where $l_a$ represents the length of the arm of the quadrotor, and since the same quadrotors are used in the flock, it is feasible to suppose that all constant coefficients are the same for all quadrotors.

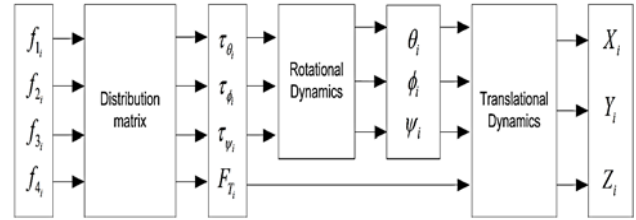

Fig.2 Scheme of the dynamics of a qaudrotor. It can be divided into Rotational dynamics and Translational dynamics.

The dynamics of a quadrotor are shown in the block diagram illustrated by Figure 2, where according to [22], the translational dynamics of a quadrotor i can be written as follows:

$$\begin{cases} \ddot{X}_i = (sin\psi_i sin\varphi_i + cos\psi_i cos\varphi_i sin\theta_i)\frac{F_{Ti}}{m} \\ \ddot{Y}_i = (cos\varphi_i sin\psi_i sin\theta_i - cos\psi_i sin\varphi_i)\frac{F_{Ti}}{m} \quad (6) \\ \ddot{Z}_i = -g + (cos\theta_i cos\varphi_i)\frac{F_{Ti}}{m} \end{cases}$$

$X_i, Y_i,$ and $Z_i$ represent the coordinates of the center of mass of the quadrotor i in the fixed inertial frame $\{e_1, e_2, e_3\}$.

$\varphi_i, \theta_i$, and $\psi_i$ as the Euler angles (Roll, Pitch and Yaw) of quadrotor i.

However, the rotational dynamics in the body-fixed frame $\{b_1, b_2, b_3\}$ is represented by: [22]

$$\begin{cases} I_{xb} \cdot \dot{p}_i + (I_{zb} - I_{yb})q_i r_i = \tau_{\varphi i} \\ I_{yb} \cdot \dot{q}_i + (I_{xb} - I_{zb})p_i r_i = \tau_{\theta i} \quad (7) \\ I_{zb} \cdot \dot{r}_i + (I_{yb} - I_{xb})p_i q_i = \tau_{\psi i} \end{cases}$$

Where scalars $I_{xb}$, $I_{yb}$ and $I_{zb}$ represent the moments of inertia of the quadrotor with respect to $x_b, y_b$, and $z_b$.

$p_i$, $q_i$, and $r_i$ represent the angular velocities of quadrotor i with respect to its body fixed frame.

Note that there exists the following relationship between the angular velocity of quadrotor i with respect to inertial frame and that with respect to body-fixed frame [22]:

$$\begin{bmatrix} \dot{\varphi}_i \\ \dot{\theta}_i \\ \dot{\psi}_i \end{bmatrix} = \begin{bmatrix} 1 & sin\varphi_i tan\theta_i & cos\varphi_i tan\theta_i \\ 0 & cos\varphi_i & -sin\varphi_i \\ 0 & sin\varphi_i/cos\theta_i & cos\varphi_i/cos\theta_i \end{bmatrix} \cdot \begin{bmatrix} pi \\ qi \\ ri \end{bmatrix} \quad (8)$$

In order to control each quadrotor, in most of the applications, the objective is to control its 3D positions, which is known as a navigation control problem. The navigation is accomplished using the desired trajectory tracking control.

The desired motion of a quadrotor can be uniquely specified by the 3D states $X_i, Y_i, Z_i$ and the yaw angle $\psi_i$. Therefore, the torques and the thrust force are calculated according to the desired moving trajectory and the desired yaw angle $\psi_i$.

It is observed from Figure 3 that the navigation control of a quadrotor has a cascade-loop property. The rotational dynamics control is in the inner loop, while the translational dynamics control is in the outer loop.

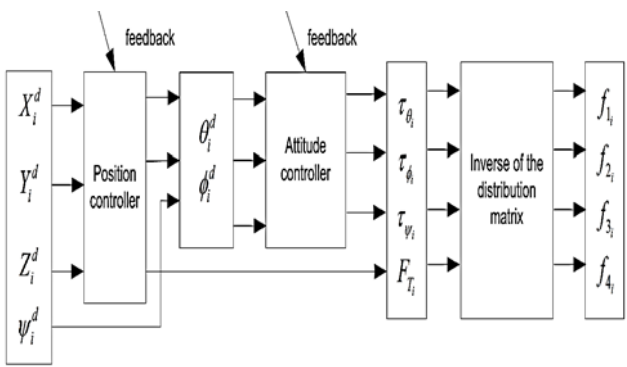

Fig.3 Quadrotor Navigation Control.

In formation flight control using Neighbor-based Trajectory Generation Topology, the trajectory will be generated for each UAV in the first place according to equation (4). So, $X_i^d$ and $Y_i^d$ in Figure 3 will be available with the given desired yaw angle and altitude $Z_i^d$. Accordingly, each UAV will be able follow its desired generated trajectory, and the group of all trajectories will form the desired formation. The formation control block diagram for each UAV is shown in Figure 4.

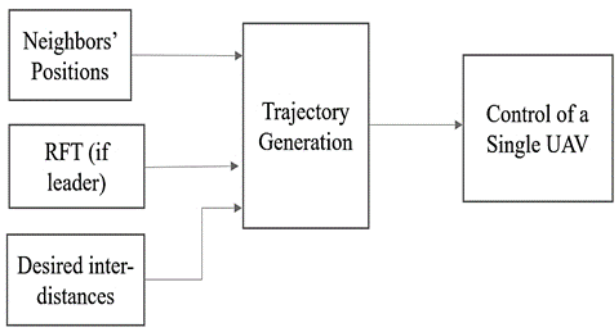

Fig.4 The block diagram of the neighbor-based trajectory generation topology. This block is for each UAV in the formation.

Note that, in this work, the objective is to apply formation flight control, where the task of the formation can be accomplished by using trajectory planning for each UAV. Because of that, the control of single agents used the flatness-based control theory but due to lack of space, it is not detailed in this paper. What is important is that all UAVs are always under control, work properly, and of course follow the given trajectory.

## 4 Simulation

To illustrate the performance of the proposed controller for Leader-Follower formation, the simulation results using MATLAB/SIMULINK is presented in this section.

In the considered example, the number of UAVs is four, $n=4$, the maximum distance of sensing is 3m, $d$ = 3m, and the initial positions of the UAVs are as follows: UAV$_1$ (0,2), UAV$_2$ (-2,0), UAV$_3$ (0,-2), and UAV$_4$ (2,0). All the UAVs has zero initial velocities. The objective of the formation is to track the circular trajectory $r(t) = [\,3sin0.1t\,,3cos0.1t\,]$ with the desired constant biases $d_{10} = [0,2]^T$, $d_{20} = [-2,0]^T$, $d_{30} = [0,-2]^T$, and $d_{40} = [2,0]^T$.

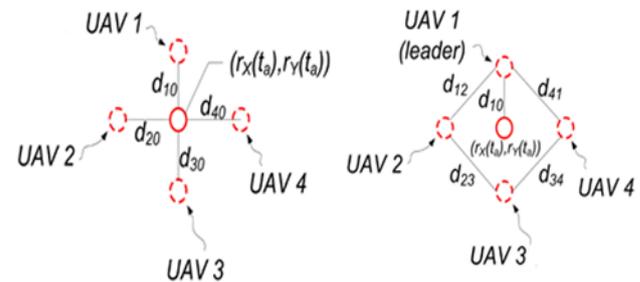

Fig.5 Formation configuration for four UAVs, where UAV$_1$ is a leader and the others are followers.

Note that the planar formation is considered, the desired altitude is $Z_{di}$ = 5m. It is worth to recall that the RFT is not available for the UAVs that are followers. However, the UAVs have the knowledge of the desired inter-distances with respect to their neighbors as shown in Figure 5. For instance, UAV$_3$ has two neighbors UAV$_2$ and UAV$_4$. The UAV$_3$ is expected to keep inter-distances $d_{32} = [2,-2]^T$ and $d_{34} = [-2,-2]$ with respect to UAV$_2$ and UAV$_4$. So, the generated trajectory will be according to (4).

It is obvious from Figure 6 that the generated trajectory is taking into account following the reference trajectory in the first place, and of course the distances between each UAV and the reference point. For example, UAV$_1$ is on the same abscissa level as the given reference, thus it is shown that its x generated trajectory is coincided with this reference. However, its ordinate has a bias of 2 units, which is why the y generated trajectory is of amplitude 5.

Thus, the four UAVs follow the generated trajectories in Figure 6 to create the formation flight control as shown in Figure 7. Accordingly, the formation in 3D is presented in Figure 8.

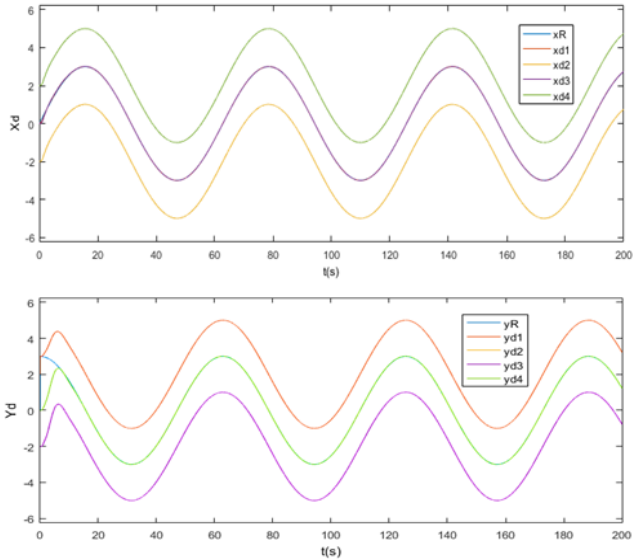

Fig.6 Generated desired trajectory of X and Y positions for each UAV in the formation.

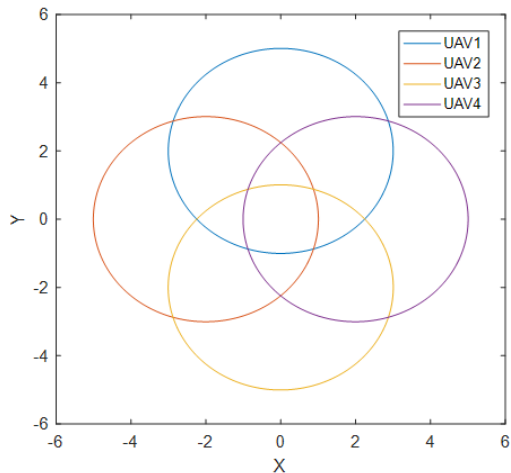

Fig.7 The Formation flight of the four UAVs using Neighbor based trajectory generation topology in 2D. Overlapping between UAVs' trajectories takes place.

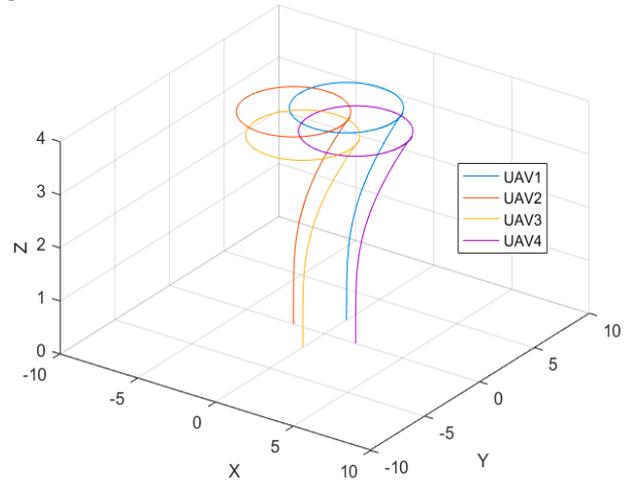

Fig.8 The Formation flight of the four UAVs using Neighbor based trajectory generation topology in 3D.

Even though overlapping between the UAVs' trajectories takes place in Figures 7 and 8, depending on the application, such case can be avoided by increasing the inter-distances or Biases, as shown in Figure 9 and 10 in 2D and 3D respectively, but one should pay attention for the sensing range.

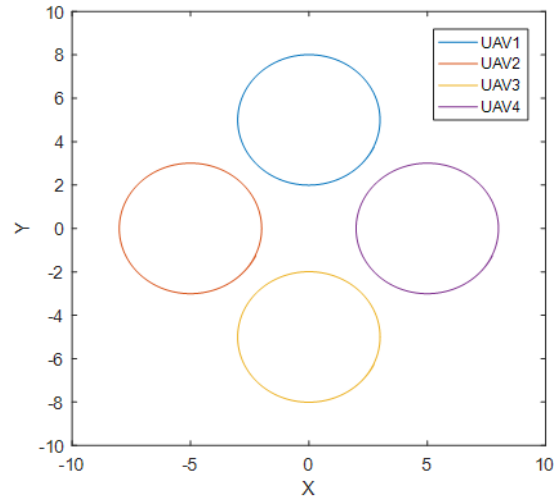

Fig.9 The Formation Flight of the four UAVs using Neighbor-based Trajectory generation Topology in 2D without trajectories overlapping.

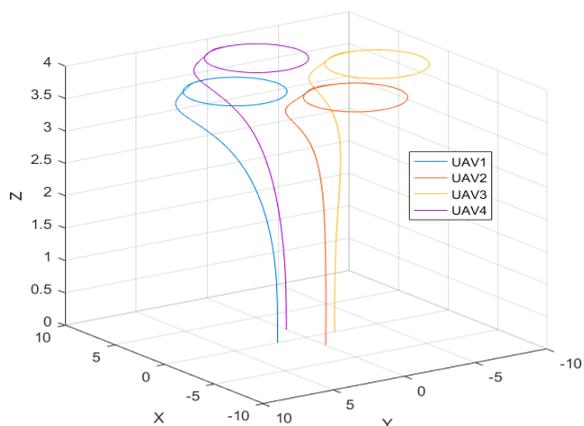

Fig.10 The Formation Flight of the four UAVs using Neighbor based Trajectory generation Topology in 3D without trajectories overlapping.

Note that both Roll and Pitch angles are in the accepted range, as shown in Figure 11.

As mentioned before, it is not necessary to take the same inter-distances in the whole flight time. For instance, after a certain time the desired inter-distance between the UAVs in the formation is increased while they kept following the circular desired trajectory, as shown in Figure 12.

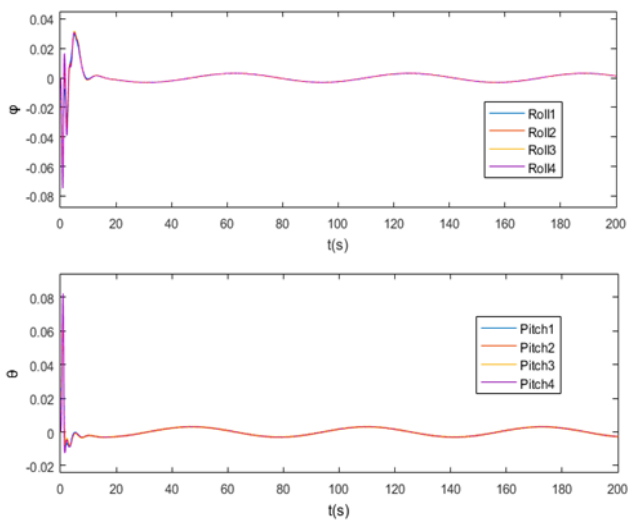

Fig.11 The Roll and Pitch angles for the UAVs in the formation. They are oscillating in the acceptable range.

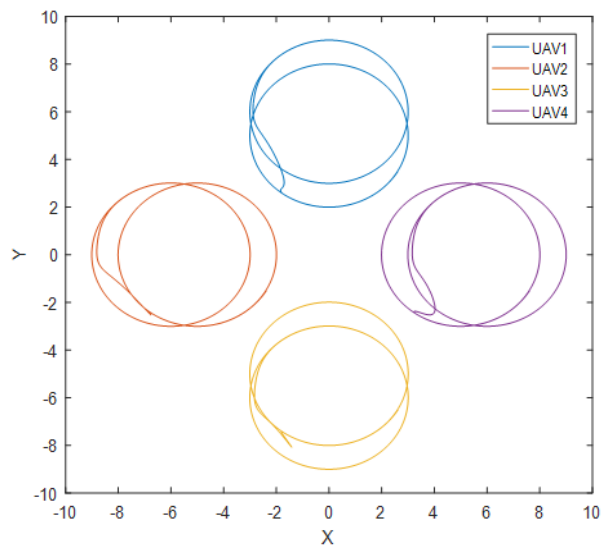

Fig.12 The Formation Flight in 2D with inter-distance change.

In order to generalize the proposed topology and to highlight the ability and simplicity of increasing the number of agents in the formation while using such a method, another example was taken into account with an increased number of UAVs, $n=6$. The maximum distance of sensing is 8.5m.

The objective of the formation is to track the same considered previous circular trajectory but with a desired constant biases forming a triangular shape: $d_{10} = [0,6]^T$, $d_{20} = [-6,0]^T$, $d_{30} = [-12,-6]^T$ $d_{40} = [0,-6]^T$, $d_{50} = [12,-6]^T$ and $d_{60} = [6,0]^T$.

Accordingly, the formation of six quadrotors was successfully achieved using neighbor-based trajectory generation topology as shown in 2D in Figure 13.

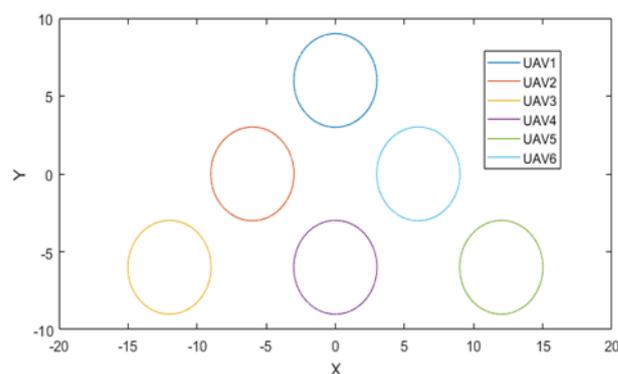

Fig.13 The Formation Flight of the six UAVs using Neighbor-based Trajectory generation Topology in 2D.

## 4 Conclusion

In this paper, a distributed Leader-Follower control method, which is the neighbor-based trajectory generation topology, was introduced for the first time, to the best of our knowledge. Where, the trajectory of each UAV in the formation is generated online taking into account following the reference trajectory given to the leader in the first place, and keeping the desired biases and inter-distances between the agents in the second place. Also, since the reference trajectory is not available for the followers, their generated trajectories take the advantage of their neighbors' positions. To illustrate such strategy, it was applied on a formation of four quadrotors, then six quadrotors, using MATLAB/SIMULINK, where the simulation has showed very good results. These results are encouraging, thus future work will consist of implementing this topology on a real experiment. In addition, this formation control approach will be compared to other techniques existing in literature.

*References:*
[1] Z. Hou, W. Wang, G. Zhang, C. Han. A survey on the formation control of multiple quadrotors. 14$^{th}$ Int. Conf. on Ubiquitous Robots and Ambient Intelligence, South Korea, 2017.
[2] K. Sreenath, T. Lee, V. Kumar. Geometric control and differential flatness of a quadrotor UAV with a cable-suspended load. IEEE 52nd Annual Conf. on Decision and Control, Italy, 2013.
[3] T. Tomic, K. Schmid, P. Lutz, A. Domel, M. Kassecker, E. Mair, I.L. Grixa, F.Ruess, M. Suppa, D. Burschka. Toward a fully autonomous UAV: Research platform for indoor and outdoor urban search and rescue. Robotics Automation Mag., IEEE, 19(3):46-56, 2012.
[4] M. Chiaramonti, F. Giulietti, G. Mengali. Formation control laws for autonomous flight vehicles. 14th Mediterranean Conference on Control and Automation, Italy, 2006.
[5] S. Montenegro, Q. Ali, N. Gageik. A review on distributed control of cooperating mini UAVs. International Journal of Artificial Intelligence & Applications (IJAIA), 5(4):1-13, 2014.
[6] J.R.T. Lawton, R.W. Beard, B.J. Young. A decentralized approach to formation maneuvers. IEEE Transactions on Robotics and Automation, 19(6):933-941, 2003.
[7] M. Saied, M. Slim, H. Mazeh, H. Shraim, C. Francis. Unmanned Aerial Vehicles Fleet Control via Artificial Bee Colony Algorithm. 4th Conf. on Control and Fault Tolerant Systems, Morocco, 2019.
[8] A. Kushleyev, D. Mellinger, C. Powers, V. Kumar. Towards a swarm of agile micro quadrotors. Aut. Robots, 35(4):287-300, 2013.
[9] A. Schollig. Synchronizing the motion of a quadrocopter to music. IEEE Int. Conf. on Robotics and Automation, USA, 2010.
[10] M. A. Guney, M. Unel. Formation Control of a Group of Micro Aerial Vehicles (MAVs). IEEE Int. Conf. on Sys., Man, and Cyb., UK, 2014.
[11] E. Semsar-Kazerooni, K. Khorasani. Optimal consensus algorithms for cooperative team of agents subject to partial information. Automatica, 44(11): 2766-2777, 2008.
[12] R. Olfati-Saber. Flocking for multi-agent dynamic systems: algorithms and theory. IEEE Transactions on Automatic Control, 51(3): 401-420, 2006.
[13] G. Antonelli. Flocking for multi-robot systems via the null-space-based behavioral control. Swarm Intelligence, 4(1): 37-56, 2010.
[14] A.S. Brandão, M. Sarcinelli-Filho. On the Guidance of Multiple UAV using a Centralized Formation Control Scheme and Delaunay Triangulation. Journal of Intelligent & Robotic Systems, 84(1-4): 397-413, 2015.
[15] A.S. Brandão, J.P.A. Barbosa, V. Mendoza, M. Sarcinelli-Filho, R. Carelli. A Multi-Layer Control Scheme for a centralized UAV formation. International Conference on Unmanned Aircraft Systems, USA, 2014.
[16] M. Hejasi, H. Noura, A. Drak. Control Theory: Perspectives, Applications and Developments, Chapter: Formation Flight of Small Scale Unmanned Aerial Vehicles: A Review. Nova Science Publishers, pp.221-248, 2015.
[17] G. Vasarhelyi. Outdoor flocking and formation flight with autonomous aerial robots. IEEE/RSJ International Conference on Intelligent Robots and Systems, USA, 2014.
[18] X. Zhang. An output feedback nonlinear decentralized controller design for multiple unmanned vehicle coordination. American Control Conference, USA, 2006.
[19] G. Vasarhelyi. Outdoor flocking and formation flight with autonomous aerial robots. IEEE/RSJ International Conference on Intelligent Robots and Systems (IROS), USA, 2014.
[20] E. Semsar-Kazerooni, K. Khorasani. Switching control of a modified leader-follower team of agents under the leader and network topological changes. Control Theory Applications, IET, 5(12): 1369-1377, 2011.
[21] Z. Hou, Isabelle Fantoni. Leader-follower formation saturated control for multiple quadrotors with switching topology. Workshop on Research, Education and Development of Unmanned Aerial Systems, Mexico, 2015.

[22] Z. Hou. Modeling and formation controller design for multi-quadrotor systems with leader-follower configuration. PhD thesis, UTC, Compiègne, France, 2016.

[23] H. Zhou, H.-L. Xiong, Y. Liu, N.-D Tan, L. Chen. Trajectory Planning Algorithm of UAV Based on System Positioning Accuracy Constraints. Electronics, 9(2): 250, 2020.

[24] A. Grancharova, E. I. Gotli, D.-T. Ho, T. A. Johansen. UAVs trajectory planning by distributed MPC under radio communication path loss constraints. Journal of Intelligent & Robotic Systems, 79(1): 115-134, 2014.

[25] V. Gonzalez, C. A. Monje, L. Moreno, C. Balaguer. UAVs mission planning with flight level constraint using Fast Marching Square Method. Robotics and Autonomous Systems, 94: 162–171, 2017.

[26] P. Ladosz, H. Oh, W. Chen. Trajectory Planning for Communication Relay Unmanned Aerial Vehicles in Urban Dynamic Environments. Journal of Intelligent & Robotic Systems, 89(1-2):7-25, 2018.

**Contribution of individual authors to the creation of a scientific article (ghostwriting policy)**

Batool Ibrahim has developed the methodology, has conducted the research, and has carried out the simulation and the analysis of the results.

Hassan Noura has guided and overviewed the work, participated in the analysis of the results, and corrected the paper.